\documentclass[12pt]{article}

\usepackage{epsfig}
\usepackage{cite}
\usepackage{amsmath, amssymb, amsfonts}
\usepackage{color}
\usepackage{xcolor}
\usepackage{latexsym}  
\usepackage{graphicx}
\usepackage{cancel}
\usepackage[colorlinks,bookmarks]{hyperref}
\hypersetup{pdfpagemode=UseNone, pdfstartview=FitH, linkcolor=blue, citecolor=red, urlcolor=blue}

\def\be{\begin{equation}}
\def\ee{\end{equation}}
\def\ba{\begin{eqnarray}}
\def\ea{\end{eqnarray}}

\def\bdm{\begin{displaymath}}
\def\edm{\end{displaymath}}
\def\la{~\mbox{\raisebox{-.6ex}{$\stackrel{<}{\sim}$}}~}
\def\ga{~\mbox{\raisebox{-.6ex}{$\stackrel{>}{\sim}$}}~}
\def\gm{\gtrsim}

\def\bq{\begin{quote}}
\def\eq{\end{quote}}

 at 10truept

\newcommand{\eps}{\epsilon}

\newcommand{\Mpl}{M_{\mathrm{Pl}}}

\newcommand{\bea}{\begin{eqnarray}}
\newcommand{\eea}{\end{eqnarray}}

\newcommand{\bi}{\begin{itemize}}
\newcommand{\ei}{\end{itemize}}

\newcommand{\beq}{\begin{equation}}
\newcommand{\eeq}{\end{equation}}
\newcommand{\beqa}{\begin{eqnarray}}
\newcommand{\eeqa}{\end{eqnarray}}
\newcommand{\mpl}{\Mpl}

\newcommand\unity{1\!\!1}

\def\12{{1 \over 2}}

\def\ltap{\ \raise.3ex\hbox{$<$\kern-.75em\lower1ex\hbox{$\sim$}}\ }
\def\gtap{\ \raise.3ex\hbox{$>$\kern-.75em\lower1ex\hbox{$\sim$}}\ }
\def\gl{\ \raise.5ex\hbox{$>$}\kern-.8em\lower.5ex\hbox{$<$}\ }
\def\roughly#1{\raise.3ex\hbox{$#1$\kern-.75em\lower1ex\hbox{$\sim$}}}

\begin{document}

\thispagestyle{empty}
\begin{flushright}
July 2026 \\
DESY 26-098
\end{flushright}
\vspace*{.35cm}
\begin{center}
{\Large \bf Wading the String Bog: CMB Birefringence in}\\
\vskip.3cm
{\Large \bf the Swampland}\\

\vspace*{.75cm} {\large Guido D'Amico$^{a, }$\footnote{\tt
damico.guido@gmail.com}, Nemanja Kaloper$^{b, }$\footnote{\tt
kaloper@physics.ucdavis.edu} and 
Alexander Westphal$^{e,}$\footnote{\tt alexander.westphal@desy.de}}\\

\vspace{.3cm} {\em $^a$Department of Mathematical, Physical and Computer Sciences,}\\
\vspace{.05cm}{\em University of Parma, 43124 Parma, Italy}\\
\vspace{.2cm}
{\em $^b$INFN Gruppo Collegato di Parma, 43124 Parma, Italy}\\
\vspace{.2cm}
{\em $^b$QMAP, Department of Physics and Astronomy, University of
California,}\\
\vspace{.05cm}{\em Davis, CA 95616, USA}\\
\vspace{.2cm} 
$^e${\em Deutsches Elektronen-Synchrotron DESY, Notkestr. 85,}\\
\vspace{.05cm}{\em 22607 Hamburg, Germany}\\

\vspace{.75cm} ABSTRACT
\end{center}
We point out that observations of cosmic birefringence may provide direct experimental 
tests of Swampland conjectures. Ultralight scalar field birefringence mechanisms are 
constrained by the Weak Gravity Conjecture, limiting their parameter space. 
Conversely, confirming that CMB birefringence is caused by ultralight scalars could spell 
problems for the Swampland framework. Resolving these difficulties without a revision of 
Swampland ideology points toward thin axionic domain walls, which can explain birefringence 
without propagating ultralight degrees of freedom. Importantly these conflicting scenarios 
for cosmic birefringence could be experimentally distinguished by a search for, or exclusion 
of, birefringence anisotropies, that could be measured by POLARBEAR, 
Simons Observatory and LiteBIRD. We also note that cosmic birefringence observables emerge
via a Sakharov-like asymmetry, where linear polarization is first generated cosmologically 
and subsequently twisted by a parity-violating dynamics after last scattering.

\vfill \setcounter{page}{0} \setcounter{footnote}{0}
\newpage

\begin{flushright}

{\it In the swamp fishing was a tragic adventure. ... There were} \\
{\it plenty of days coming when he could fish in the swamp. ~~~~~ }\\
\vskip.25cm
\hfil{\it ~~(Ernest Hemingway, ``Big Two-Hearted River")}
\end{flushright}
\vskip.5cm

\section{Introduction}

About $70\%$ of the cosmic contents is dark energy. Maybe it is a cosmological constant, but 
explaining how it would be as small as needed is challenging. Evolving alternatives such as 
quintessence are at least as challenging, if not more. For both, the requisite dynamics needs 
ultra-small scales and mechanisms which neutralize the quantum corrections, protecting these 
small numbers. In general, one needs to suppress many operators in the effective field theory (EFT) 
of dark energy which is generically very hard \cite{Wilczek:1983as,Weinberg:1988cp,Polchinski:2006gy}.

Even if some mechanism gets rid of vacuum energy, the problems do not go away. 
If observations \cite{DESI:2024mwx,Kamionkowski:2022pkx} converge on evolving 
dark energy, we'd need a consistent theory of quintessence. If it is a field theory, it must 
have a very shallow potential, which should be quantum-mechanically stable
to be natural. Perturbative stability in EFT can be enforced with approximate symmetries, 
but when non-perturbative and/or quantum gravity
corrections are included, these symmetries are lifted and the problems are typically aggravated.  

The issue is that to behave as dark energy, field theories usually 
require flat potential plateaus 
to support large field variations  of order ${\cal O}(\mpl)$. 
There may be problems with such large field variations because they may 
require energies which may backreact strongly on the background. 
These concerns are linked to the weak gravity conjecture \cite{Arkani-Hamed:2006emk}, which is based on
the current lore about quantum gravity (QG), 
and subsequent additional conjectures discussed within 
the framework of the ``Swampland" program \cite{Ooguri:2018wrx,Agrawal:2018own,Palti:2019pca}. 
The critical question is whether field theories which model such large field ranges can emerge from
UV completions that include QG. Direct observations of dynamical 
dark energy might therefore provide an arena to test some ideas
about Planck scale physics, and ultimately QG. Given 
the scarcity of experimental tests of anything relating to QG, 
this offers an exciting opportunity to consider, 
and begs for additional means to test this regime. 

Fortuitously, the issue appears even in the simplest limits of 
quantum field theory (QFT) coupled to
semiclassical gravity. For example, we may argue that the observational hints for
dynamical dark energy  \cite{DESI:2024mwx,Kamionkowski:2022pkx} 
combined with the need to maintain control over EFT of dark energy favor 
axion-like degrees of freedom (hereafter, axions) as a very attractive
quintessence candidate
\cite{Frieman:1995pm,Nomura:2000yk,Kim:2002tq,Kaloper:2008qs,DAmico:2018mnx,Ibe:2018ffn,Kaloper:2025goq}. 
Yet, such models typically also require field variations $\sim {\cal O}(\mpl)$. Hence they too 
appear to require reckoning with gravitational backreaction, which very interestingly
encroaches on the domain of QG \cite{Nicolis:2008wh}. This apparent tension, which is often viewed as a 
challenge,  could be a blessing 
in disguise.  

It turns out that there is a different class of observationally accessible signals 
which could be sensitive to QG. If dark energy is an axion-like quintessence, it could also 
couple directly to electromagnetism via
the electromagnetic Chern--Simons term,  
\be
{\cal L} \supset \frac{g}{4} \phi F_{\mu\nu}\tilde F^{\mu\nu}\, ,
\label{axpho}
\ee
where $\tilde F^{\mu\nu} = \frac12 \epsilon^{\mu\nu\lambda\sigma} F_{\lambda\sigma}$. 
This coupling induces cosmic birefringence of light in a background which involves a varying axion 
\cite{Komatsu:2022nvu,Huang:1985tt,Harari:1992ea,Carroll:1998zi,Lue:1998mq,Pospelov:2008gg,
Minami:2020odp,Choi:2021aze,Diego-Palazuelos:2022dsq,Gasparotto:2022uqo,Ferreira:2023jbu}
because the two photon helicities are affected
differently by axion variation. This happens because the 
Chern--Simons term is parity-odd. As a result, the evolution of the
cosmic microwave background (CMB) through a spatial region occupied by a varying axion field rotates the
polarizations of different helicities at a different rate, imprinting a signal in the CMB. There are tantalizing hints
that this signal might be nonzero, being of order of $0.3$ degrees of angle, or, in radians,
\be
\Delta \vartheta \sim 10^{-3} \, ,
\ee
giving the variation of the polarization of the CMB between the last scattering and now. 

In what follows, we will see that if this signal is due to the time-varying ultralight axion quintessence, then this field
should change by a lot, possibly by as much as $\ga {\cal O}(\mpl)$ in order 
to explain this signal naturally. Therefore confirming that CMB polarization rotation is
not zero could corroborate the suspicions that some fields in the universe do change by more than
Planckian shifts. Further, since such slow variations are accumulated 
along the (very long!) line of sight, they are generically anisotropic, 
where the rotation angle may vary by ${\cal O}(1)$ factors 
from direction to direction. Finally, they could affect other sources of light
generated after last scattering, such as radio galaxies, providing additional checkpoints for the presence of 
slowly rolling axion quintessence.

There is, however, an alternative explanation of CMB birefringence which does not involve ultralight axions 
\cite{Kaloper:2025goq,Kaloper:2026slg,Kaloper:2026gib,Kaloper:2026ygk}. 
As it turns out, in the event that the leading order polarization 
rotation of the CMB is isotropic, the 
natural interpretation would be that it is caused by CMB crossing very thin axion domain walls. Those walls are
naturally associated with the recent Discretely Evanescent Dark Energy proposal \cite{Kaloper:2025goq}. In this case,
the CMB polarization rotation is imprinted instantly every time CMB light crosses a domain wall \cite{Kaloper:2026ygk}.
Hence the total polarization rotation merely depends on the final and initial vacuum. Generically, the former 
is the same for all light rays arriving to our telescopes by default, and the latter is prepared by causal
early universe dynamics such as, e.g. inflation, and so the accumulated polarization rotation is the same in all directions. 

Interestingly, the appearance of a measurable birefringence signal in the CMB requires 
two distinct steps. First, cosmological perturbations and Thomson scattering generate 
a linearly polarized CMB after recombination. Second, this polarization must encounter 
a sector which violates parity and twists the polarization plane. Hence observable 
cosmic birefringence follows from dynamics closely related to 
Sakharov's mechanism for baryogenesis \cite{Sakharov:1967dj}. An asymmetry must first be generated, 
and then processed by interactions which violate the relevant symmetries. 
We will return to this perspective below, because it provides an intuitive picture for both 
ultralight-axion and vacuum-interface birefringence.

Crucially, observational tests may discriminate between these outcomes, which are 
within reach of cosmological experiments such as POLARBEAR, Simons Observatory and 
LiteBIRD. Thus, if CMB birefringence were to be confirmed, it could be a rare {\it additional} check of ideas about
quantum gravity. The outcome where birefringence is isotropic can coexist with Swampland ideas, since 
it does not require large field excursions. Conversely, anisotropic birefringence would point toward
ultralight fields with large field variations, which may challenge Swampland ideology. We
remain agnostic about the outcome since both options are very interesting.

\section{Adiabatic CMB Birefringence and Ultralight Axions}

Nonzero polarization rotation induced in the CMB while it propagates from the 
last scattering surface  at the epoch of recombination to us suggests that there is a 
nonzero contribution 
$\propto \langle \vec E \cdot \vec B \rangle \propto \langle F_{\mu\nu}\tilde F^{\mu\nu}\rangle$ 
in the CMB polarization field, where $\langle \ldots \rangle$ denotes averaging over the sky. 
Since this term is CP-odd, it would be natural to attribute its presence, if it is
confirmed, to a dynamics which involves an axion-like degree of freedom in some guise (hereafter an ``axion"), 
with a coupling to electromagnetism given by \eqref{axpho}. Until recently the ideas about how 
to produce such a signal invariably involved ultralight axions, which vary sufficiently slowly to yield an 
adiabatic twist of the CMB polarization plane 
\cite{Huang:1985tt,Harari:1992ea,Carroll:1998zi,Lue:1998mq,Pospelov:2008gg,
Minami:2020odp,Choi:2021aze,Diego-Palazuelos:2022dsq,Gasparotto:2022uqo,Ferreira:2023jbu,Fedderke:2019ajk}. 

The logic leading to ultralight axions is not subtle. 
CMB is generated very late in the universe, around recombination, when neutral atoms are
formed, after which the universe becomes mostly transparent. Hence the mechanism
which induces birefringence should turn on very late. If an axion is very light,
it will be frozen by cosmic expansion as long as $m < H$, at $\phi \sim f$, and so \eqref{axpho}
will be an irrelevant total derivative for most of the cosmic history. Only after the expansion
rate dips below axion mass will the axion field start varying, which is necessary 
for \eqref{axpho} to become non-negligible. Truncating the axion potential to its mass term, an
initially displaced axion will obey $\ddot\phi + 3H\dot\phi + m^2\phi \simeq 0$. 
Since recombination occurs during matter domination, with $H > H_0$,  
the axion is a subleading spectator at that time, and its equation admits the exact solution
$\phi(t) = {\tt a}\,f t_i {\cos[m(t-t_i)]}/{t}$ where ${\tt a}$ is a ${\cal O}(1)$ number that  
parameterizes initial condition $\phi \sim f$; $t_i$ is the time $\phi$ begins to oscillate 
and $f$ is an axion decay constant. In the adiabatic regime 
the CMB polarization rotation induced by such an evolution in adiabatic approximation
where $m < \omega_{\tt CMB}$ is bounded by 
(see, e.g. \cite{Kaloper:2026slg,Kaloper:2026gib,Kaloper:2026ygk} 
for a complete analysis of various regimes of birefringence)
\be
\Delta\vartheta \simeq \frac{g \Delta \phi}{2} 
\lesssim \frac{g f}{2} \, {\tt a} mt_i \int_{t_i}^{t_f} \!dt\,\frac{\sin[m(t-t_i)]}{t} \, .
\label{riemannleb}
\ee
In the interval $\Delta t = t_f - t_i \lesssim H^{-1}$, the factor
$1/t$ decreases slowly, while $\cos(mt)$ oscillates rapidly when $m\gg H$.
By the Riemann--Lebesgue lemma the contributions to the integral will be suppressed except inside
an interval of a single period around the initial time,
yielding at most $\Delta\vartheta \propto {\cal C} {\tt a} mt_i/2mt_i \simeq {\cal C} {\tt a}/2$, where
${\cal C} = g f$. 

Hence in this approach if $m> H_{\tt LSS}$ this will occur before the CMB is around. Even if $\phi$ is still
oscillating after the last scattering surface, the imprint on the CMB will be too suppressed by the premature 
release of the axion $\phi$. 
To maximize the imprint we need $m \lesssim H$, where $H \lesssim H_{\tt LSS}$. 
Up to ${\cal O}(1)$ numerical factors, this requires
\be
m \lesssim 10^{-28}\, {\rm eV} \, ,
\label{massl}
\ee
to ensure the chances for an observable effect.
Moreover, any such signal depends sensitively on the history of the
axion evolution and can be contaminated by inhomogeneities in geometry, matter
distribution, or local dynamics. 

\section{Flavors of Axion Quintessence in the Swampland}

Let us first briefly review the slow roll quintessence models, with particular attention to the 
models involving axions. We will focus on the scaled-down models
of slow roll monodromy inflation
\cite{Silverstein:2008sg,Kaloper:2008fb,Dong:2010in,Kaloper:2011jz,McAllister:2014mpa,DAmico:2017cda}
since they are well defined UV-complete QFTs in good shape
concerning the chances for UV completions. We will not review the theoretical foundation of these
models, but merely focus on their phenomenological applications 
as dark energy. For this, the key ingredients are flat wide potentials 
and initial conditions where the field is originally far from the minimum. In this case, 
the slow roll equations lead to the equations governing the homogeneous
dark energy component, with the equation of state $w = p/\rho$,
\be
\eps = \frac{\mpl^2}{2}\left(\frac{V'(\phi)}{V(\phi)}\right)^2\, , \qquad \eta = \mpl^2  \frac{V''}{V}\, ,
\qquad w \simeq -1 + \frac23 \eps \, , \qquad \dot w \simeq \frac{4}{3} H\eps \Bigl(\eta - 2 \epsilon\Bigr) \, .
\label{slowroll}
\ee
Here primes are field derivatives and dots time derivatives. 
We will consider scalars with the effective Lagrangian
\be
{\cal L} = -\frac12 (\partial \phi)^2 - V(\phi)\, ,
\label{scallag}
\ee
where we can use the approximate power law potentials $V = \mu^4(\phi/f)^p$ 
when the attraction basins are shallow and wide,
with our choice of parameterization of $V$ becoming clear momentarily. Here
$\mu$ is the scale 
of dark energy, 
$\mu \sim 10^{-3} {\rm eV}$, and the axion decay constant $f$ is a measure of the width of 
potential attraction basin around the closest minimum. The Eqs. \eqref{slowroll} reduce to
\ba
&&\eps = \frac{p^2}{2} \left(\frac{\mpl}{\phi}\right)^2\, , \qquad \qquad \qquad ~~\,
\eta = (p-1)p \left(\frac{\mpl}{\phi}\right)^2\, , \nonumber \\
&& w \simeq -1 + \frac{p^2}{3} \left(\frac{\mpl}{\phi}\right)^2 \, , 
\qquad \qquad \dot w \simeq -\frac{2}{3} H p^3 \left(\frac{\mpl}{\phi}\right)^4 \, .
\label{slowrolls}
\ea
Here $H \sim H_0 \sim 10^{-33} ~{\rm eV}$ is the present expansion rate of the universe. 
As we will see shortly, these equations are very useful to understand what it takes for the quintessence
model to realistically fit the current dark energy. 

From the standpoint of quantum field theory, realizing such dynamics is highly nontrivial. 
The scalar must be extraordinarily light, with mass $m^2 \sim \partial^2_\phi V \lesssim H_0^2$ 
to still be in slow roll. Its potential must remain stable under radiative corrections; 
a generic scalar coupled to heavier degrees of freedom receives corrections 
$\Delta m^2 \sim \frac{g^2 m_{\rm heavy}^4}{\mpl^2}$ which will 
overwhelm $H_0^2$ unless the coupling $g$ is extremely suppressed. 
At the same time, if the scalar is so light, its exchange may mediate long-range 
forces comparable to gravity unless its couplings are very small. 
These requirements imply that any viable quintessence field must be highly protected, 
essentially decoupled from heavy sectors, and endowed with a symmetry that suppresses quantum corrections 
and long range forces between matter particles. This strongly favors pseudoscalars with 
approximate shift symmetries, since a continuous shift symmetry forbids a 
potential at the perturbative level and allows only nonperturbative contributions. For these reasons,
CP-odd fields fit the role of elusive dark energy very well \cite{DAmico:2016jbm}. 

The simplest realization comes from some axion coupled to a dark sector, with a 
potential arising from a dark gauge field instanton effects, 
which in the dilute instanton gas approximation give
\cite{Frieman:1995pm}
\be
V(\phi) = \mu^4 \left(1 - \cos\frac{\phi}{f}\right)\, .
\ee
The scale $\mu$ must be $\sim 10^{-3} \, {\rm eV}$ to fit dark energy now. The potential width must be
$f \gm \mpl$ to ensure enough slow roll. While these choices are clearly special, 
such models are technically natural, and do not yield long range forces.
The small scale $\mu$ is protected because in the limit $\mu \to 0$ the continuous shift 
symmetry $\phi \rightarrow \phi + \phi_0$ is restored, and the long range forces are suppressed 
since only derivative couplings to matter are allowed.

To study cosmological evolution toward the minimum, 
we can truncate the potential to the quadratic term when $f > \mpl$, 
$V(\phi) \simeq \frac12 m^2 \phi^2$ where $m^2 = \frac{\mu^4}{f^2}$. Plugging this into formulas \eqref{slowrolls}  
gives $w \simeq -1 + \frac{4}{3} \left(\frac{\mpl}{\phi}\right)^2$ and 
$\dot w \simeq -\frac{16}{3} H \left(\frac{\mpl}{\phi}\right)^4$. Clearly, to realize quintessence-like behavior, simulating the
vacuum energy for an age of the universe, we need $w+1$ and $\dot w/H$ to be roughly 
within at most a few percent. This requires $\phi \ga {\cal O}(10) \, \mpl$. 

The required field range can be reduced if we take a potential which is flatter than the quadratic. In the setups 
studied in \cite{Dong:2010in,McAllister:2014mpa,DAmico:2017cda,DAmico:2018mnx}, the mechanisms which can 
flatten such potentials were outlined, amounting essentially to taking the low energy quintessence theory near its
UV cutoff, while staying within the regime of validity of EFT. In those cases, the potentials tend to 
depend more weakly 
on the field ranges. Focusing on the flattened power potentials and ignoring the higher derivative kinetic corrections
we can see that if e.g. we employ potentials such as $V \sim \phi^{1/10}$, we can shorten the required field range
dramatically. Indeed, from \eqref{slowrolls} we see that the field range 
can be reduced by a factor of ten or so, easily. 
That takes us to the territory where to have axion-like monodromy quintessence we'd need 
barely $\sim {\cal O}(\mpl)$ field ranges. And yet, as we will see below, 
this may still be a step too deep in the bog. Fairly generic arguments \cite{Arkani-Hamed:2006emk,Nicolis:2008wh}
already indicate that even the ${\cal O}(1) \, \mpl$ field displacements involve energies which can trigger gravitational 
backreaction. We will give this a close look after introducing dark energy birefringence mechanisms. 

\section{Ultralight Birefringence Model Building}

The obvious and immediate questions are, why would such ultralight dark energy pseudoscalars couple to
electromagnetism via \eqref{axpho}, and would the coupling, if nonzero, be sufficiently large 
to induce the rotation angle $\vartheta \sim 10^{-3}$. These questions are related, and to understand them we briefly recall
how the axion-photon coupling \eqref{axpho} arises in the first place.
Recall that the ``original" axion arises as the pseudo--Nambu--Goldstone boson of a global 
$U(1)_{\rm PQ}$ symmetry that is spontaneously broken at a scale $f$ 
\cite{Peccei:1977hh,Weinberg:1977ma,Wilczek:1977pj}. 
This symmetry is explicitly broken by quantum effects, most importantly through 
the QCD anomaly, which generates the effective coupling 
$\sim \frac{\phi}{f}\,\frac{\alpha_{\tt s}}{8\pi}\,Tr(G_{\mu\nu}\tilde G^{\mu\nu})$,
that shifts $\theta_{\tt QCD}$ by the axion {\it vev}, and enables the axion
to minimize strong CP-breaking phase once nonperturbative QCD effects induce a potential for $\phi$.

In addition, other gauge interactions, and specifically electromagnetic interaction, contribute to the Peccei--Quinn (PQ)
anomaly  via mixed anomaly terms \cite{Kim:1979if,Shifman:1979if,Zhitnitsky:1980tq,Dine:1981rt}. This happens
because the fermions charged under PQ also carry electromagnetic 
charge, which mixes the PQ anomaly with $U(1)_{\rm EM}$. 
After integrating out the massive charged fermions, this yields an effective 
interaction between the axion and the electromagnetic field,
\be
\mathcal{L} \supset  C_\gamma \,\frac{\alpha_{\tt QED}}{8\pi}\,  \frac{\phi}{f} \,F_{\mu\nu}\tilde F^{\mu\nu} \, ,
\label{axionem}
\ee
where $C_\gamma$ is a dimensionless coefficient determined by the model-dependent charge 
assignments of the underlying theory. Generically it is ${\cal O}(1)$. 
The axion decay constant $f$ can take a range of
values, but importantly for many string constructions 
it is ${\cal O}(M_{\tt GUT})$. Thus, the effective QCD axion-photon coupling is
\be
g_{\phi \gamma \gamma}  =  C_\gamma \,\frac{\alpha_{\tt QED}}{2\pi f} \, .
\label{axphoto}
\ee

However, since the QCD axion's mass is 
$m_{\tt QCD} \sim (\Lambda_{\tt QCD}^2)/f \sim 10^{-10}$--$10^{-4} \, {\rm eV}$,
the QCD axion is too heavy to induce CMB birefringence in the adiabatic limit.\footnote{There are also suggestions  
to employ lighter axions for the strong CP problem, with masses outside of the usual axion window 
\cite{Hook:2018jle,DiLuzio:2021pxd}. From the birefringence phenomenology 
viewpoint, however, they do not behave very differently than more 
common axions, and so we will ignore this distinction.} 
While it would satisfy $m < \omega_{\tt CMB}$,
it would not stay in slow roll until after recombination 
and last scattering. Thus, if one wants to induce cosmic birefringence in the
CMB this way, one must communicate this 
coupling to a dark ultralight axion, which must be very light, 
much lighter than the QCD axion. This is not a problem in principle: 
due to nonperturbative 
effects in quantum gravity and gauge theory, it is harder to keep 
visible and dark sectors completely decoupled than to couple them. 
The light axions will certainly mix by some irrelevant operators, which will arise from 
integrating out heavy states and/or nonperturbative dynamics.

The question is, will those mixings be sufficiently large? From the model-building viewpoint, 
since when using the ultralights, the polarization rotation affects
CMB at scales $\omega_{\tt CMB} \sim 10^{-4} \, {\rm eV}$, we need to
communicate the axion-photon coupling to the dark sector by 
mixing it with the visible sector at the level of relevant and marginal operators. That
implies using kinetic and mass mixings. To illustrate the possibilities and the obstacles 
we first consider the simplest toy model where QCD axion and 
ultralight axion mix by terms like
\be
\mathcal L = \frac{1}{2} (\partial_\mu \phi_{\tt QCD} , \partial_\mu \phi)
\begin{pmatrix}
1 & \kappa \\
\kappa & 1
\end{pmatrix}
\begin{pmatrix}
\partial^\mu \phi_{\tt QCD} \\
\partial^\mu \phi
\end{pmatrix} 
- \frac{1}{2} (\phi_{\tt QCD} , \phi)
\begin{pmatrix}
m_{\tt QCD}^2 & \mu^2 \\
\mu^2 & m^2
\end{pmatrix}
\begin{pmatrix}
\phi_{\tt QCD} \\ \phi
\end{pmatrix} + \frac14 g_{\phi \gamma \gamma} \, \phi_{\tt QCD} \,F_{\mu\nu}\tilde F^{\mu\nu} \, .
\label{quadmix}
\ee

Before diagonalizing only $\phi_{\tt QCD}$ couples to the electromagnetic Chern--Simons term. 
The mixing communicates the coupling to the other mode. The mixing also splits the two modes further, by
level repulsion. However, an analogue of the level repulsion in the kinetic terms makes one of the new
propagation eigenmodes couple more strongly, eventually turning into a ghost \cite{Holdom:1985ag}. 
Likewise, the mass mixing decreases the mass of
the lighter mode and eventually induces a tachyonic instability. Hence $\kappa$ and $\mu^2$ are both 
bounded from above to preserve perturbativity of the low energy theory. By inspecting \eqref{quadmix},
those bounds are, respectively, 
\be
 |\kappa| < 1  \, , \qquad \qquad \mu^2 < m m_{\tt QCD} \, .
 \label{mixbounds}
 \ee

In the limit $m \ll m_{\tt QCD}$ relevant for the physical 
application to ultralight-axion induced birefringence, imposing 
the constraints \eqref{mixbounds} to simplify eigenmasses and Chern--Simons couplings, 
we can re-diagonalize the system to new propagation eigenstates. The limit is subtle,
since the problem involves three mass scales, and we have a hierarchy $m < \mu < m_{\tt QCD}$. 
With some wisdom after the fact, we choose to explore the regime  
where $\varepsilon = 1-\mu^4/(m^2_{\tt QCD}\, m^2)$ is tiny: this makes the lighter 
of the two axions parametrically much lighter relative to the input scales. 
In this regime, 
\be
m^2_- \;\simeq\; \varepsilon \, m^2 \, , \quad g_- \;\simeq\;
g_{\phi \gamma \gamma} \;
\frac{m}{m_{\tt QCD}} \, , 
\quad
m^2_+ \;\simeq\; \frac{m_{\tt QCD}^2}{1-\kappa^2} \, , \quad
g_+ \;\simeq\;
\frac{g_{\phi \gamma \gamma}}{\sqrt{1-\kappa^2}} \, .
\label{mgterms}
\ee
From these simple formulas, 
we see that mixing will spread around the Chern--Simons coupling from the QCD sector. However
the coupling strengths are normalized by $m/m_{\tt QCD}$.
If we were in the regime where $\varepsilon < {\cal O}(1)$, staying away from the tachyon boundary, 
and the dark axion is ultralight with $m \sim 10^{-33} \, {\rm eV}$, while 
the QCD axion mass is in the range $10^{-10}$--$10^{-4} \, {\rm eV}$, the coupling will be 
nonzero, but generically tiny. Cranking up the heavy axion 
coupling by kinetic mixing $\sim 1/\sqrt{1-\kappa^2}$ is possible,
but not very helpful since it does not affect the lighter coupling, cranks up the heavy mass,
and so it actually pushes down the lighter coupling $g_-$ relative to $g_+$. For these reasons we will  
steer clear from the strong coupling regime 
(see e.g. \cite{DAmico:2018mnx} and references therein), and set the 
kinetic mixing involving ultralight axions aside for the most part, ignoring scale enhancements beyond ${\cal O}(1)$.

However $\varepsilon$ in \eqref{mgterms} signals another way to link QCD 
to the dark sector. If $\varepsilon \ll 1$, and $m < m_{\tt QCD}$ is not too small, 
one can end up with an ultralight axion with mass $m_- \sim \varepsilon^{1/2} m$, whose 
coupling to the electromagnetic Chern--Simons might not be supremely small since 
$g_-/g_+ \simeq m/m_{\tt QCD}$. Thus, if e.g. $m/m_{\tt QCD} \sim 10^{-5}$, we could end up with an
interesting set of parameters $g_- \sim 10^{-18} \, {\rm GeV}^{-1}$ and $g_+ \sim 10^{-13} \, {\rm GeV}^{-1}$ and 
the ultralight axion mass as low as $m_- \sim 10^{-33} \, {\rm eV}$. Of course, in the case of this variant of the
\underline{\it see-saw} mechanism for mass hierarchy generation, we'd need a tiny $\varepsilon \sim 10^{-28}$ or so, which 
looks disturbingly out of place. 

This problem is familiar from general sweeps for ultralight axions in axion landscapes
\cite{Kaloper:2008qs,Bachlechner:2019vcb,Demirtas:2021gsq}. The general construction process 
can be assisted with the inclusion of additional axion directions. 
Specifically, we can extend the scales-inducing 
see-saw to include $N$ fields using the guise of ``Clockwork" \cite{Kaplan:2015fuy}, in order 
to induce a tiny ultralight mass while not suppressing the axion-photon couplings to minuscule 
values -- but as we will see this comes with a price. 
Let us unpack briefly the miens of the ``Clockwork" approach that
pertain to our goals. The axion mixings studied in \cite{Kaplan:2015fuy} are claimed to yield two
results: 
\begin{itemize}
\item induce a ``ladder" of mass hierarchies between initially degenerate axion fields;
\item suppress the lighter axion couplings by large ratios, so that their field ranges are
correspondingly extended.
\end{itemize}
For our purposes here, the former result is greatly beneficial, since it can be used to
explain the presence of the ultralight axion needed to drive late cosmic acceleration. However
the latter result would be detrimental if it were generic, since it would greatly suppress 
the ultralight's coupling to the photon, and prevent it from inducing desired degree of polarization
rotation. 

Serendipitously, the latter result found in \cite{Kaplan:2015fuy} is not generic. It followed from 
the implicit but definitive assumption that only one of the axions in the degenerate subspace 
before the symmetry breaking which induces the ``Clockwork descent" of axion masses is 
coupled to the gauge sector where the instantons inducing the axions' potentials originate. 
In \cite{Kaplan:2015fuy} this choice is deliberate since this is needed to ensure that the 
lightest axion has a larger field range. That's the price they elect to pay. 

Here, however, we need a coupling to the photon which is considerably less suppressed. Thus we can take 
a system of axions in which one generic linear combination 
couples to the QCD Chern--Simons $Tr(G\tilde G)$ and, by extension due to the PQ charges, 
to the electromagnetic Chern--Simons density $F\tilde F$. If we let some of 
these axions be approximately degenerate with the QCD axion, $m_i \sim m_{\tt QCD}$, 
the mixing angles can be ${\cal O}(1)$, ``virializing" the electromagnetic coupling among 
the propagation eigenstates, $g_i \sim {\cal O}(\frac{m}{m_{\tt QCD}} g_{\phi \gamma \gamma})/\sqrt{N}$. 
This is easy to understand because couplings are components of a 
vector in the axion isospace, and mass mixings are $SO(N)$ rotations
which do not change its total length. Hence for some 
generic orientation, where all the components are
comparable, their individual size is given by 
${\cal O}(\frac{m}{m_{\tt QCD}}g_{\phi \gamma \gamma})/\sqrt{N}$, by Pythagoras' theorem.

In order to avoid problems with strong CP restoration based on the QCD axion, and exclude the additional
axions from contributing (and possibly misaligning) the QCD minimum for $\theta_{\tt QCD}$, we either
need to align the QCD axion with a single protected direction in field space, or to resort to 
a two--stage construction. In the latter case, we can first introduce
a sector of $N$ axions $\phi_i$ with comparable masses,  
lighter than the QCD axion scale $m < m_{\tt QCD}$, but not ultralight, and simultaneously define a 
single QCD direction $\vec k = \{k_i\}$ 
\be
\phi_{\tt QCD} = \sum_{i=1}^N k_i\, \phi_i \, , 
\label{QCDax}
\ee
such that only this combination couples to $Tr(G\tilde G)$. Then we can arrange the mass matrix so 
that the $\phi_{\tt QCD}$ direction is an eigenvector, $M^2 \vec k \propto \vec k$. 
This direction will not mix with others, and so the QCD coupling is blocked from the modes orthogonal to
\eqref{QCDax}. On the other hand, by choosing the charges of the electromagnetic sector we 
can still communicate an ${\cal O}(g_{\phi \gamma \gamma})$ 
electromagnetic coupling of the QCD axion to the light dark 
sector without introducing extremely strong parametric suppression.

In the second stage, we finally induce an internal hierarchy 
within this dark axion subsector by adding a ``Clockwork"-style
mass matrix
\be
\mathcal L_{\rm mass}
=
-\frac12 m^2 \sum_{i=0}^{N-1} (\phi_i - q\,\phi_{i+1})^2 \, ,
\label{axmasses}
\ee
where $q > 1$. We can pick $q \simeq 10$ which rapidly generates the needed mass 
hierarchy without requiring an excessive effort to justify it.
This will only happen after the breaking of symmetry which induces \eqref{axmasses}.
Crucially, since the electromagnetic coupling is already distributed across the sector,
\be
\mathcal L \supset \frac{g_{\phi \gamma \gamma}}{4} \left(\sum_i c_i \phi_i\right) F\tilde F,
\qquad \qquad c_i \sim \frac{1}{\sqrt{N}} \frac{m}{m_{\tt QCD}} \, ,
\ee
the diagonalization of the ``Clockwork" sector can produce ultralight 
eigenstates without further parametric suppression of the coupling. In this way we can obtain ultralight axions with
\be
g_{\rm ultra} \sim \frac{g_{\phi \gamma \gamma}}{\sqrt{N}}  \frac{m}{m_{\tt QCD}} \, ,
\label{ultracoup}
\ee
instead of the much smaller value $\sim g_{\phi \gamma \gamma}\, m_{\rm ultra}/m_{\tt QCD}$ that 
would arise from direct, single step mixing. The key point 
is that coupling redistribution and hierarchy generation are separated
relative to the original proposal in \cite{Kaplan:2015fuy}: since mass mixing alone is inefficient for coupling 
transfer, we communicate these couplings before invoking ``Clockwork" mechanisms to 
generate large hierarchies.

Importantly, in the construction outlined above, the electromagnetic 
coupling is distributed across the chain of the ``Clockwork" cogs. This avoids inducing 
an additional exponential suppression of the coupling. Further, this coupling ``spreading" may be
more generic, and hence robust, after the inclusion of additional corrections 
which may also involve gravity-induced universal terms
that do not stay confined to a specific subsector but are ``flavor democratic". 

This is more aligned with the Swampland ideology. 
Achieving subsectors which are sequestered from each 
other is generically harder, either because of universal perturbative couplings to
the graviton multiplet, gravitational nonperturbative sectors or 
challenges with stabilizing internal dimensions and brane configurations 
embedded in them. Note that the original ``Clockwork" which realizes very large field ranges \cite{Kaplan:2015fuy} does 
so by following a deconstruction-like prescription \cite{Arkani-Hamed:2001kyx}
where only a single axion couples to a gauge sector
before the clock starts ticking. This is perfectly reasonable in QFT, but given the outcome and the possibly super-Planckian
field ranges which result, it may be contentious in the Swampland 
framework according to the current lore\footnote{An alternative philosophy was pursued
in \cite{Nomura:2000yk,Ibe:2018ffn}, where the idea is to imagine a phase like the strong-CP violating $\theta_{\tt QCD}$
in some UV extension of the Standard Model, mix it with the electroweak sector, and introduce an axion-like
degree of freedom which couples to the electroweak Chern-Simons and picks an extremely small mass term
from electroweak instantons after the corresponding global symmetries are broken. However this axion
must be extremely well protected from all other UV effects, even more than the modes above.}. 

To get the idea about the parameter range that could yield
interesting phenomenology,  let us take $q=10$. Then the lightest mass is suppressed by a 
factor $10^{-N}$ relative to $m$. If the 
intermediate dark sector has a characteristic scale $m \sim 10^{-15}\,\mathrm{eV}$, not too far from the
QCD axion mass range, to obtain an ultralight mode with 
$m_{\rm ultra} \sim 10^{-33}\,\mathrm{eV}$, one requires\footnote{Strictly speaking, this
would also imply that there are other ultralights with masses within a few orders of 
magnitude of $m_{\tt ultra}$, which could also affect both background geometry and birefringence, if they
have comparable couplings. Such additional ultralight fields could be avoided if instead of field-theoretic ``Clockwork" one
employs a multi-flux supported see-saw, generalizing the example of \cite{Kaloper:2008qs}.}%
\be
\frac{m_{\rm ultra}}{m} \sim 10^{-18} \;\;\Longrightarrow\;\; N \sim 18 \quad (q=10) \, .
\label{counting}
\ee
Thus a chain of order $N \sim 15$--$20$ axions is sufficient to generate 
an enormous hierarchy of masses without explicitly introducing tiny parameters. In this 
regime, the lightest mode remains parametrically light while inheriting an electromagnetic coupling of order
$g_{\rm ultra} \sim \frac{g_{\phi \gamma \gamma}}{\sqrt{N}}  \frac{m}{m_{\tt QCD}}$, 
provided the coupling vector is democratically distributed across the sites, as we noted above.  
This would realize the desired separation between hierarchy generation 
(via ``Clockwork") and coupling strength (set at the intermediate stage), yielding 
ultralight axions with phenomenologically relevant electromagnetic interactions.

We stress that this does not guarantee that the first-principles constructions do 
yield field theories where so many axions automatically mix with each other and the
required axion-U(1) couplings. In many known cases, fewer fields are involved.
On the other hand, ubiquity of axion ``quality" problems and the universality of gravity
seem to obstruct excluding such options at this time. Hence we feel that more work is needed to
resolve how likely such multi-axion mixings are. 

In any case, using \eqref{axphoto} and the bounds on axion-photon couplings, with 
$m \sim 10^{-15} \, {\rm eV}$ and $m_{\tt QCD} \sim 10^{-10} \, {\rm eV}$ and
$q = 10$ and $N = 18$, we now get $g_{\tt ultra} \sim {\rm few} \times 10^{-19} \, {\rm GeV}^{-1}$,
when the QCD axion's coupling to electromagnetism is close to the bounds, 
$g_{\phi\gamma\gamma} \sim 10^{-13} \, {\rm GeV}^{-1}$. The ultralight axion mass 
is $m_{\tt ultra} \sim 10^{-33} \, {\rm eV}$, so it is a slow roll field until today, more or less. 
Of course, our estimates are somewhat optimistic: for example, a stronger future
bound on the axion-photon coupling or on the QCD axion mass could easily limit the ultralight's coupling
to electromagnetism more. Nevertheless, at least at the present moment, as we are about to explain below,
the current limits may meet the phenomenological requirements to fit the
claimed hints of CMB birefringence. 
 
\section{Ultralight Birefringence Phenomenology}

For a given axion profile $\phi$, in the absence of charges photon dynamics
reduces to standard axion electrodynamics with field equations (see \cite{Kaloper:2026ygk} for a 
comprehensive analysis)
\be
\partial_\mu \Bigl\{ F^{\mu\nu} + \frac{g}{2}   
\epsilon^{\mu\nu\lambda\sigma}  \phi F_{\lambda\sigma} \Bigl\} = 0 \, .
\label{mmonax}
\ee
Here we are working in a locally Lorentzian frame, ignoring cosmic expansion, which can
be restored by minimal metric coupling prescription, restoring redshift dependence of the
non-scale invariant quantities. We set all charges and currents to zero. 
Substituting $F_{\mu\nu} = \partial_\mu A_\nu - \partial_\nu A_\mu$ the Maxwell 
equations \eqref{mmonax} in Lorentz gauge reduce to
\be
\partial^2 A^\mu =
g \epsilon^{\mu\nu\lambda\sigma}
\bigl( \partial_\nu \phi \bigr) \partial_\lambda A_\sigma \, . 
\label{mmonaxpot}
\ee
Using residual gauge freedom in Lorentz gauge
we set $A^\mu = (0,\vec A_{\perp},0)$, where $\vec A_{\perp}$ is the
transverse vector field satisfying
$\vec\nabla \cdot \vec A_{\perp} = 0$. Substituting this into
Eq.~\eqref{mmonaxpot} and Fourier transforming in (conformal) time,
$\vec A_{\perp} \propto e^{-i\omega t}$, after straightforward algebra\footnote{We can safely ignore the subtlety that
involves the tachyonic instability of horizon-size wavelength vector modes, which arises when
$\partial_\mu \phi$ is a timelike vector \cite{Campbell:1992hc,Anber:2009ua} because we are working with 
much shorter wavelengths here.} 
we find
\be
\Bigl(\bigl(\frac{d}{dz}\bigr)^2 + \omega^2 \Bigr)
\begin{pmatrix}
A^x_\perp \\
A^y_\perp
\end{pmatrix} =
g \omega \partial_z \phi
\begin{pmatrix}
0 & -i \\
 i & 0
\end{pmatrix}
\begin{pmatrix}
A^x_\perp \\
A^y_\perp
\end{pmatrix} \, .
\label{mateq}
\ee
We stress that this equation holds both for a very large domain wall and for an evolving 
scalar field background along the line of sight, since the time coordinate and the 
``radial" coordinate $z$ are related by $\int dt = \int dz$. We also stress that despite this 
one should not conflate the cosmologically varying ultralight axions with very light axion
``hair" sourced by thick domain walls. These are very different phenomena, and while in
the former case the control of birefringence is exerted by the ultralight's initial conditions 
as illustrated by Eq. \eqref{riemannleb}, for thick domain walls the polarization rotation
is set by the wall's thickness, which is {\it still} controlled by the topology of axion vacua. 
While formulas are formally similar, their physical interpretation is completely different.  

In the adiabatic regime, $\omega \gg g\,\partial_z\phi$, the medium is not 
absorptive and propagating photons transmit unitarily. We may take
all photons to propagate along characteristics defined by 
$i\partial_z-\omega=0$, corresponding to the look-back time prescription.
In this case, projecting onto ingoing modes only, 
$\Bigl(\bigl(\tfrac{d}{dz}\bigr)^2+\omega^2\Bigr) 
\simeq 2\omega(\omega-i\partial_z)$, we reduce 
Eq.~\eqref{mateq} to a Schr\"odinger equation with
``imaginary time'' $iz$,
\be
i \partial_z
\begin{pmatrix}
A^x_\perp \\
A^y_\perp
\end{pmatrix} =
H
\begin{pmatrix}
A^x_\perp \\
A^y_\perp
\end{pmatrix} \, , \qquad \qquad 
H = \omega \unity - \frac{1}{2} g \partial_z \phi \, \sigma_2 \, ,
\label{nonmateq}
\ee
where $\sigma_2$ is a Pauli matrix. Mapping wave propagation to a first-order
equation in this way is standard in analyses of axion--photon propagation
\cite{Raffelt:1987im}.

Solving Eq.~\eqref{nonmateq} is straightforward because the
Hamiltonians at different ``times'' $z$ commute. 
When $\omega \gg |g\,\partial_z\phi/2|$, the off-diagonal term in $H$ can be
treated as a perturbation. Defining 
$\Delta \phi = \int_{-\infty}^{\infty}dz\,\partial_z\phi$, we find 
$\int_{z_0}^{z}dz\,H = \omega \Delta z \unity
+ \frac{1}{2} g \Delta\phi \sigma_2$, and so finally 
\be
\begin{pmatrix}
A^x_\perp(z) \\
A^y_\perp(z)
\end{pmatrix}
=
e^{-i\omega \Delta z}
\begin{pmatrix}
\cos(\frac{1}{2} g \Delta \phi) & -\sin(\frac{1}{2} g \Delta \phi) \\
\sin(\frac{1}{2} g \Delta \phi) & \cos(\frac{1}{2} g \Delta \phi)
\end{pmatrix}
\begin{pmatrix}
A^x_\perp(z_0) \\
A^y_\perp(z_0)
\end{pmatrix} .
\label{rotsoln}
\ee
Thus for wavelengths
$\lambda\ll 1/m_\phi$, the polarization vector undergoes adiabatic rotation by
an angle
\be
\vartheta = \frac12 \, g \, \Delta\phi \, ,
\label{polrot}
\ee
relative to its initial orientation. This angle measures the relative phase accumulated by the 
two circular polarization states of the photon as they propagate through cosmological spacetime.

This is easily confirmed using the wave dispersion relation \cite{Harari:1992ea}. From the wave equation 
the two circular polarizations propagated by \eqref{mateq} obey $\omega_\pm^2 = k^2 \pm g k \dot\phi$
where we traded the look-back distance for time, and so the average net rotation arises from the
propagation time delay for waves moving in the direction $\hat n$, after solving the dispersion relation
for $\omega_\pm$, 
\be
\vartheta(\hat n) = \frac{1}{2}\int_\gamma dt\, \big(\omega_+ - \omega_-\big) 
\simeq  \frac{1}{2}\int_\gamma dt\, k 
\Bigl((1+ \frac{g \dot \phi}{2k}) - (1- \frac{g \dot \phi}{2k})\Bigr) = \frac12 {g \Delta \phi} \,.
\ee
As noted this derivation is valid in the adiabatic regime, where the time variation of $\phi$ is slow 
compared to the photon frequency. In the non-adiabatic regime, where axion variation is fast, 
one must solve the equations more carefully, although 
the result remains formally a sum over phase shifts induced by discontinuities 
of the effective ``vacuum angle" $\theta \sim  g \Delta \phi$ when reflection from the regions with
axion gradients is negligible. This occurs to leading order whenever 
$g \Delta \phi \ll 1$ \cite{Kaloper:2026ygk}. 

We can now check the numbers that it takes to fit the magnitude of the observational hint of 
nontrivial cosmic birefringence. Per \cite{Komatsu:2022nvu}, the polarization rotation angle, if 
confirmed, may be in the range of $\vartheta \simeq 0.3$ degrees, or 
\be
\vartheta \sim 5 \times 10^{-3} \, {\rm radians} \, .
\label{radprec}
\ee
The optimistic estimate of the ultralight's coupling
to electromagnetism is $g_{\rm ultra} \sim \frac{g_{\phi \gamma \gamma}}{\sqrt{N}} \frac{m}{m_{\tt QCD}}$ 
noted in the previous subsection, which with our choice of ``Clockwork" parameters 
$N \sim 18$, $m \sim 10^{-15} \, {\rm eV}$ and $m_{\tt QCD} \sim 10^{-10} \, {\rm eV}$ 
as well as $g_{\phi \gamma \gamma} \simeq 10^{-13} \, {\rm GeV}^{-1}$ yields 
\be
g_{\tt ultra} \sim {\rm few} \times 10^{-19} \, {\rm GeV}^{-1} \, . 
\label{ultrapheno}
\ee
With the relation of the axion decay constant and the axion-photon coupling in Eq. \eqref{axphoto} 
this implies that $f \sim 10^{11} \, {\rm GeV}$.  
Alternatively, if the fundamental axion decay constant is higher, $f \sim 10^{16} \, {\rm GeV}$, as
often encountered in string constructions, with all other parameters being the same we find 
\be
g_{\tt ultra} \sim {\rm few} \times 10^{-24} \, {\rm GeV}^{-1} \, . 
\label{ultraphenomin}
\ee
Therefore the polarization rotation angle induced by the dark energy ultralight's variation in the CMB, using 
\eqref{polrot}, is  
\be
\vartheta = \frac12 \, g \, {\cal O}(\mpl) = 10^{-5} - 0.1 \, \, {\rm radians} \, .
\label{totrads}
\ee
Clearly some of the wiggle room in \eqref{totrads} may shrink over time, both due to the theory input and
from observational improvements. 
 
\section{Dark Energy Birefringence in the Swampland}

It is evident from the examples above that explaining CMB birefringence with varying
CP-odd dark energy fields encroaches on the realm of QG. We need both large field ranges and 
very weak couplings, that both appear to be set by some physics that ``manufactures" the Planck scale. 
Here we finally turn to the implications of WGC \cite{Arkani-Hamed:2006emk} pertaining
to axions. To start we consider a simple case of a single axion which receives a non-perturbative potential from 
instanton effects, with the aim of using it as a theory of dark energy. 
This would correspond to the examples of \cite{Frieman:1995pm,Nomura:2000yk,Kim:2002tq}. The 
conjecture asserts that there exists an instanton of action $S$ satisfying the so called ``electric" WGC bound 
$f S \lesssim \mpl$.\footnote{There is still ongoing discussion in the literature as to 
which `strength'~\cite{Arkani-Hamed:2006emk} the WGC should ultimately have. 
Variants range from the mild version demanding only the existence of at least one 
state satisfying the conjecture to the much stronger lattice/sublattice 
\cite{Heidenreich:2015nta,Heidenreich:2016aqi,Montero:2016tif} and 
tower \cite{Andriolo:2018lvp} versions. For a review, see \cite{Harlow:2022ich}. 
We will work with the milder version as the most defensible version, which can be
easily supported by the bottom-up arguments \cite{Arkani-Hamed:2006emk}.} 
This implies that the potential induced by this instanton generates a contribution to the potential,
$V(\phi) \sim \Lambda^4 e^{-S} \cos(\frac{\phi}{f})$, which corrects the axion mass by 
$\delta m^2 \sim V''(\phi) \sim \frac{\Lambda^4}{f^2} e^{-S} \ga \frac{\Lambda^4}{f^2} e^{-\mpl/f}$. 
If $f \ga \mpl$ to allow at least marginal slow roll, $S < {\cal O}(1)$, 
and the instanton effects lose exponential suppression. In fact this 
was already noted before the advent of WGC based on the generic 
behavior of the instanton expansion in string theory in \cite{Banks:2003sx}. 
While the first correction to the axion mass is $\delta m^2 \sim \Lambda^4/f^2$, which can be ${\cal O}(H_0^2)$ iff  
$\Lambda \sim 10^{-3} \, {\rm eV}$ is the dark energy scale, the higher order corrections are not negligible, and 
must be resummed before deciding if the theory can be a dark energy candidate or not. Alternatively, if
one picks $S \gg 1$, one ends up with $f \ll \mpl$, and the theory will not have a broad flat plateau. 

Thus to make the axion field light enough and flat enough to be quintessence, one runs into tension 
with the conjecture. If $f \ga \mpl$, to satisfy the conjecture, one needs to accommodate additional 
instantons with smaller action or allow for light states. Both of these conditions spoil the simple 
ultralight EFT description. By going to the strong coupling regime, 
using flattening from irrelevant operators getting large, one might find a plateau even when the fundamental decay 
constant is close to $\mpl$. However, to ensure this one needs full control over  an infinite series of irrelevant 
operators organized by specific symmetries, such as the discrete gauge symmetry of a 
$3$-form dual description. Without it one cannot be certain that a generic plateau introduced 
by hand would be radiatively stable and under perturbative control. It is possible to get assistance from the ${\cal O}(100)$
numerical factors in the loop expansion as argued in \cite{DAmico:2017cda,DAmico:2018mnx}, 
but one still needs to explain where the very small axion masses come from. 

Since in our specific worked out example of ``Clockwork"-like portal for communicating the electromagnetic
Chern--Simons term to the ultralight dark sector we relied on many light fields, we can explore these more
intricate details. Because the ultralight axion gets its mass and coupling to $F \tilde F$ from mixing with see-saw
rather than direct nonperturbative effects, the straightforward application of the ``electric" WGC bound used above 
is more obscure.
A more practical approach is to use the magnetic version of WGC, which was discussed 
in \cite{Arkani-Hamed:2006emk,Hebecker:2015zss}. There, it was shown that the QG arguments derived from 
WGC reasoning yield an upper bound on the cutoff
of the sector of the theory that produces the nonperturbative potential that originally breaks the relevant
PQ symmetry. Basically the WGC bound can be imposed on the spherical domain walls which separate
the nontrivial vacua of the gauge theory, which supplies the instanton-induced vacuum energy
that lifts the UV vacuum degeneracy after confinement. Introducing an axion which relaxes to the CP-restoring
minimum energy vacuum, and monodromizing it with the flux sourced by the gauge theory domain walls
relates the WGC bound for the domain walls to the axion  periodicity so that the discrete axion shift 
symmetry remains unbroken. The WGC bound relates the domain wall tension and coupling to the 
Planck scale, ${\cal T} < g \mpl$, and since for a monodromized axion $g \sim mf$, this yields the bound
$\Lambda^3 \sim mf \mpl$ \cite{Hebecker:2015zss}, where $\Lambda$ is the EFT cutoff.

In our case, this comes from the QCD axion, which, in order to solve the strong CP problem needs 
the effective cutoff $\Lambda_{\tt QCD} \simeq {\rm few} \times 100 \, {\rm MeV}$. Because in our
example, our ultralight axions mix with the QCD one to pick up coupling to the electromagnetic Chern--Simons term,
and because they must not break the residual discrete shift symmetries of the axion vacuum manifold in the IR,
the bound constrains $mf$ of all the axion fields which 
inherit the coupling and their decay constant by
\be
\Lambda_{\tt QCD}^3 \la m f \mpl \, ,
\label{magcut}
\ee
where we assume generic mixings as discussed above. 
Therefore, the axion masses which come from this sector should be bounded by, roughly, 
\be
 m \ga \frac{\Lambda_{\tt QCD}^3}{\mpl \, f} \sim \frac{10^{-21} \, {\rm GeV}^2}{ f} \, .
\label{axmassmagcut}
\ee
Given the possible values of $f$ for the axion from the previous section, $f \sim 10^{11}$ -- $10^{16} \, {\rm GeV}$,
we see that the magnetic WGC bound for an axion places the bounds on the ultralight mass $m$ in the range
\be
m \ga 10^{-28} \, {\rm eV} \, - \, 10^{-23} \, {\rm eV} \, .
\label{masswiondow}
\ee
Remarkably, this is in the right ballpark to signal a direct tension with Eq. \eqref{massl} which shows how light an axion
must be to have a chance of inducing the right amount of CMB polarization rotation. Since 
our arguments here are imprecise, as we did not reproduce carefully all of the required numerical
coefficients, we won't go as far as claiming an exclusion of the ultralight axion induced CMB birefringence
producing polarization rotation by an angle in \eqref{radprec}.\footnote{See, however, 
recent progress~\cite{DiUbaldo:2026rly,Maldacena:2026jqd} using arguments 
based on the exact results for the gravitational path integral of axionic (Giddings-Strominger type) 
wormholes implying a determination of the precise numerical coefficient 
for the `electric' WGC in 4D for axions: $fS < \frac12 \pi\sqrt{3}\mpl$.} 
However, it should be clear that there
is a tension between the Swampland bounds and invoking an ultralight axion to explain 
polarization rotation by $\vartheta \sim 10^{-3}$ radians. 

In fact the physical potential for this conflict can be seen with a ``naked eye". To have the ultralight be quintessence
we need $f > \mpl$ to make sure the late domination by dark energy lasts long enough. On the other hand, 
since $g \sim 1/f$, increasing $f$ decreases the polarization rotation. Thus the hint that $\vartheta$ might be
a promille of radian pushes in the direction opposite to the requirements for dark energy. Ergo, the ultralight explanation
of CMB birefringence can be ``risky". 

\section{Land Ho!}

Given the tension found above, a natural question is if the positive confirmation of CMB birefringence
forces a direct conflict between such observations and the Swampland ideology. Interestingly, in the light
of the very recent work  \cite{Kaloper:2026slg,Kaloper:2026gib,Kaloper:2026ygk} the answer is in the negative. 
The traditional approach of the past  four decades to inducing CMB birefringence by 
insisting on the adiabatic approximation, which relies on ultralight axions, is not the only game 
in town any more. Instead, it is perfectly viable to go beyond the adiabatic approximation
and use axions much heavier than quintessence, to the point that they can be integrated out altogether from the low
energy theory \cite{Kaloper:2026slg,Kaloper:2026gib,Kaloper:2026ygk}. The only requirements are that the low energy theory
has a nontrivial system of vacua with different topological properties, separated by domain walls which bear the residual
Chern--Simons couplings to electromagnetism which remain when the axions are integrated out. 

When the axion is decoupled, so that $\partial\phi \rightarrow 0$, it becomes 
frozen in one of these vacua. The resulting low energy effective action describing the
electromagnetic waves propagating through a system of such vacua is 
\be
S = -\frac{1}{4}\int d^4x\, F_{\mu\nu} \, F^{\mu\nu}
- \frac{1}{4}\int d^4x\, \theta(x)\, F^{\mu\nu} \, \tilde F_{\mu\nu} \, .
\label{action}
\ee
Where $\theta$ is constant, 
\eqref{action} is just conventional electrodynamics. In the approximation of infinitely thin walls between 
regions with different values of $\theta$, we can model the walls by writing 
$\theta(z) = \theta_- + \Delta\theta \, \Theta(z)$. Since the jump $\Delta\theta$ at $z=0$ 
induces a Chern--Simons interaction on the interface,
the electromagnetic waves crossing it get ``twisted" in a helicity-dependent way.
The precise and detailed analysis of this process is given 
in \cite{Kaloper:2026slg,Kaloper:2026gib,Kaloper:2026ygk}.
We merely summarize its principal features here. 

The replacement of the axion-photon coupling $g \phi \rightarrow \theta$ with the domain wall in 
\eqref{axpho} leads to the substitution $g \partial_z \phi \rightarrow \Delta \theta \delta(z)$, 
where we approximate the wall with a plane, which is valid in the limit when the
wall's curvature radius is much larger than the wavelength of the photons crossing the wall. The latter
is on the order of a millimeter for the CMB and shorter for more localized sources that we may also 
consider. Hence this is not an issue. The field equation \eqref{mateq} thus reduces to 
\be
\Bigl(\bigl(\frac{d}{dz}\bigr)^2 + \omega^2 \Bigr)\vec A_\perp
= \Delta \theta \,\omega\,
\delta(z)\,
\sigma_2\,\vec A_\perp\, .
\label{mateqsm}
\ee
We can solve this equation exactly. Inverting it, we rewrite it as 
a completely equivalent $1+1$-dimensional Lippmann-Schwinger scattering equation, 
with the exact solution \cite{Kaloper:2026ygk}
\be
\vec A_\perp(z) =
e^{-i\omega z}\,\vec A_\perp(+\infty) - i\,\frac{\Delta \theta}{2}\,\sigma_2 \,
\vec A_\perp(0)\,e^{i\omega |z|}\, . 
\label{borneqs}
\ee
Solving for 
$\vec A_\perp(0)$ in terms of $\vec A_\perp(+\infty)$, we find $\vec A_\perp(0) = \frac{1}{\sqrt{1+\hat \sigma^2}} 
e^{- i\,\vartheta\,\sigma_2} \vec A_\perp(+\infty)$, where we use shorthand
notation $\hat \sigma = \Delta \theta/2$ and $\vartheta = \tan^{-1}(\hat \sigma)$.
Substituting into the solution \eqref{borneqs} yields 
the exact solution of \eqref{mateqsm} describing an incoming
wave refracting on the wall,
\be
\vec A_\perp(z) =
\Bigl(e^{-i\omega z}\, \unity - i\,\sin(\vartheta)\, e^{i\omega |z|} \, 
e^{- i\,\vartheta\,\sigma_2} \, \sigma_2  \Bigr) \vec A_\perp(+\infty) \, .
\label{borneqsa}
\ee
This describes both the transmitted wave ($z<0$) and the incident+reflected wave ($z>0$). 

The polarization orientation changes discretely at the thin wall 
for all wavelengths of photons below the cutoff of the theory
describing the wall at low energies. This precisely matches 
the Pancharatnam phase definition 
\cite{Pancharatnam:1956url,Pancharatnam:1956url2}. Using the
helicity eigenbasis for the photon, $A_\pm(z)
= \frac{1}{\sqrt{2}}\Bigl(A_\perp^x(z) \pm i A_\perp^y(z)\Bigr)$, the 
transmitted waves in the region $z<0$ are 
\be
A_\pm(z<0) = \cos(\vartheta)\, e^{\mp i\vartheta}\,
e^{-i\omega z}\, A_\pm(+\infty),
\qquad
A_\mp(z<0)=0.
\label{transcirc}
\ee
The two circular polarizations cross the wall independently from each other, and pick up 
helicity-dependent phase factor $e^{\mp i\vartheta}$. The Pancharatnam phase is 
the relative phase between the complex amplitudes,
\be
e^{i\gamma_{\rm P}^{(\pm)}}
=
\frac{A_\pm(+\infty)^\ast\,\tilde A_\pm(z<0)}
{|A_\pm(+\infty)^\ast\,\tilde A_\pm(z<0)|}
=
e^{\mp i\vartheta} \, .
\label{panchph}
\ee
Hence the Pancharatnam phase acquired by the two helicity components is
\be 
\gamma_{\rm P}^{(\pm)} = \mp\,\vartheta \simeq \mp \frac{\Delta \theta}{2} \, ,
\label{panfinal}
\ee
in the limit of small $\Delta \theta$. In fact, the natural value of $\Delta \theta$ 
can be extracted from \eqref{action} and our definition of the couplings, to yield
\be
\Delta \theta = g \Delta \phi \simeq g f \simeq C_\gamma \frac{\alpha_{\tt QED}}{2\pi} 
\simeq 10^{-3} \, C_\gamma \, .
\label{estrot}
\ee
Thus a $C_\gamma \sim {\cal O}(1)$ could well fit the hints that $\vartheta \simeq 10^{-3}$ 
even when the axion is completely absent from the low energy theory. All it takes is that
the theory retains the memory of the axion vacua, and that the propagating photons 
traverse different domains separated by domain walls bearing electromagnetic Chern--Simons terms.

The only remaining question is, why would the CMB photons cross such domain walls so late
in the universe. A natural answer is provided by the recent proposal for Discretely Evanescent
Dark Energy \cite{Kaloper:2025goq}. Briefly, this proposal entails a dark sector which has a non-Abelian 
gauge group with dark quarks, which goes strong like QCD, but at a much lower scale. With minor arranging 
of the theory's matter contents, the strong coupling scale can be as low as $10^{-3}\, {\rm eV}$. 
When none of the dark quarks are massless after the theory confines, the chiral symmetry breaking 
condensate provides the vacuum energy $\sim 10^{-12} \, {\rm eV}^4$. We can reinterpret the condensate
as the flux of a ``secret top-form" originally introduced in QCD by L\"uscher \cite{Luscher:1978rn} 
(see also e.g. \cite{Gabadadze:1997kj,Gabadadze:2002ff}), which can discharge by the emission
of membranes with charges and tensions set by the dark sector strong coupling scale. 

Remarkably, the decay rate and the lifetime are easily on the order of the present age of the universe,
and further the decay rate is $\Gamma \sim H_0^4$, which means that when the universe reaches
the present age, the production of the dark membranes becomes very prolific. The bubbles are produced so 
fast that they easily percolate \cite{Guth:1982pn,Turner:1992tz}, stopping cosmic acceleration after that 
time and rendering dark energy a transient phenomenon. 
\begin{figure}[bth]
    \centering
    \includegraphics[width=12.5cm]{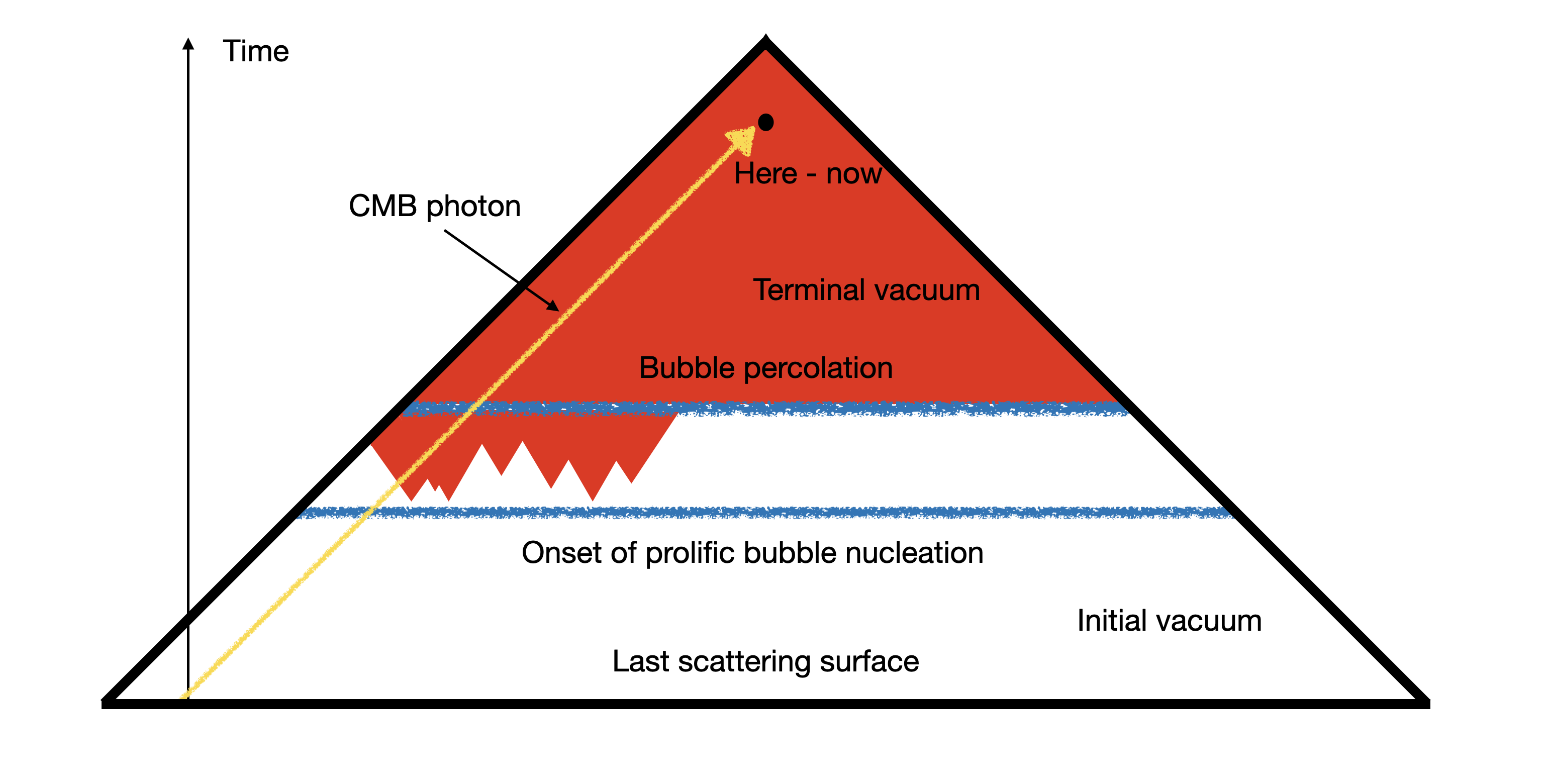}
    \caption{A CMB photon arriving from the last scattering surface.}
    \label{fig1}
\end{figure}

As noted in \cite{Kaloper:2025goq,Kaloper:2026slg,Kaloper:2026gib,Kaloper:2026ygk}, the dark walls
naturally support the electromagnetic Chern--Simons terms. Those couplings at low energies can arise either
from the mixing of the dark top-form with the visible sector axion/top-form, or by mixing of the visible
and dark axions along the lines described above -- with the exception that the transmission of the coupling
may be simpler since no ultralight axions are ever involved. Further, this also ensures that in this scenario
there will be no Planckian field ranges traversed, and hence there will be little if any friction with the
Swampland ideology. Finally, the charges and tensions of the emergent membranes which play the
role of the domain walls between vacua easily satisfy the WGC constraints for extended 
objects  \cite{Arkani-Hamed:2006emk,Hebecker:2015zss}.

Now we can consider a theory with a discrete set of vacua parameterized 
by $\theta$, exactly degenerate prior to a late-time, post-last scattering symmetry breaking, that could
explain late-emerging dark energy as in \cite{Kaloper:2025goq}. Taking the early universe to originate from
a single causal domain (e.g. by being prepared by early inflation), we have the same initial vacuum 
across the entire observable patch, fixing the value $\theta_{\rm initial}$ everywhere. Subsequently 
a dark confining phase transition lifts the degeneracy, induces dark energy, and enables vacuum decay
to proceed via bubble nucleations. Domain walls separate regions with different values 
of $\theta$. Assuming that the decay process begins after recombination, and ends before the 
present epoch, in accordance with \cite{Kaloper:2025goq}, $\theta$ is locally constant within 
each domain and changes only across walls. For illustration see Fig. \ref{fig1}.

Then along any photon trajectory $\gamma$, the field $\theta$ jumps only at the wall crossings.
Entering a bubble, $\theta$ jumps by a discrete amount $\Delta\theta_i$; if the photon departs the bubble, then
$\theta$ jumps by $-\Delta\theta_i$. Since all CMB photons originate from the last scattering surface,
where they are all in the same initial vacuum, and all are observed in the same final vacuum here and now,
we can rewrite the sum over all crossings, $\vartheta(\hat n) = \sum_i \Delta\theta_i$ (where the sign of the jump in
$\theta$ may flip depending on the orientation of the wall crossing with respect to the wall outward normal)
as a path integral 
\be
\vartheta(\hat n) = \int_\gamma d\theta \, ,
\label{path}
\ee
since $d\theta$ vanishes everywhere except at the walls. Because $\theta$ 
labels vacua and no strings are present, the map between spacetime and vacuum 
space is single-valued, which implies that $d\theta$ is globally exact: 
for any closed contour $C$, $\oint_C d\theta = 0$. Hence, for open contours the 
path integral \eqref{path} depends only on the endpoints of $\gamma$. 

Explicitly,
$\vartheta(\hat n) = \theta(\text{endpoint}) - \theta(\text{starting point})$. 
With our boundary conditions being the same along any line of sight, this implies 
\be
\vartheta(\hat n) = \theta_{\rm final} - \theta_{\rm initial} = \text{constant}\, ,
\ee
independent of the direction of the line of sight $\hat n$. This means that wall-induced 
birefringence signal is exactly isotropic, which can be tested observationally \cite{Lonappan:2025hwz}. 
The role of domain walls is thus to mediate transitions between vacua, supplying 
discrete contributions to $\vartheta$ along photon paths. However, because the total 
change in $\vartheta$ between the endpoints is fixed, the number and arrangement of
walls along the line of sight do not affect the result. Further, since the low energy
effective action \eqref{action} is manifestly conformally invariant \cite{Kaloper:2026ygk}, 
the accumulated polarization
rotation angle will not depend on the distance to the walls even after we transition
to full FRW metric. The final result is redshift-independent. 
The quantity $\vartheta(\hat n)$ 
counting crossings of walls of different orientation reduces to a topological invariant 
determined solely by the difference of $\theta$ between initial and final vacua. 

Anisotropy could be generated if 1) $\theta$ is a field, and inflation does not homogenize its initial state 
2) the dark energy decay begins before last scattering or continues until the present 
in a tuned way, different lines of sight probe different initial or final vacua, or 3) if dark cosmic 
strings are present, so that the integral acquires path dependence. Further  
if $\theta$ is promoted to an ultralight axion, it may fluctuate copiously and 
one may recover the bulk scenario with anisotropies. Each of these modifications introduces 
additional structure not present in the minimal framework and moves the theory 
toward the class already constrained by the WGC analysis. The distinction between these 
two mechanisms is therefore sharp. Bulk ultralight fields necessarily produce 
anisotropic birefringence because fluctuations of the field 
imprint direction-dependent contributions along photon paths. 
Domain wall induced birefringence without ultralight axions yields 
a strictly isotropic signal fixed by the difference between initial 
and final vacua. These outcomes are therefore mutually exclusive and in the sights of CMB
polarization measurements in the near future.

\section{Begetting the Chiral Sky}

Now we turn to the microscopic picture which describes how observable signatures of
birefringence arise in the CMB. This is a problem which has been studied extensively for ultralight axions
in adiabatic approximation, both foundationally \cite{Lue:1998mq} and with primary focus on practical
aspects of extracting the signals from the data, as in, e.g. \cite{Minami:2020odp,Nakatsuka:2022epj} and many
other references. Here we will focus on the thin walls and birefringence beyond the adiabatic
approximation, since this is a new approach to the problem, and in fact it turns out to be
remarkably simple physically. This simplicity could already be glimpsed from the phase accumulation
description in the previous section, which shows that the dynamics of generating polarization
rotation is ultralocal, in contrast to being spread around over large optical depth as in the 
case of ultralight axions. 

To understand what happens, we find the insightful clues in  \cite{Lue:1998mq} particularly useful.
There the authors note a similarity of the generation of birefringent signal in the CMB to the emergence of 
baryon asymmetry of the universe. They note that since this asymmetry is imprinted in the photon
background, which is a bath of chargeless states, one requires parity and time reversal breaking
by both the local dynamics {\it and} the background in order to generate a nontrivial signal. 

In fact this is a remarkable observation, which leads us to the fact that the conditions for generating 
an observable parity-violating signature in the CMB are a variant of Sakharov's conditions for 
baryon asymmetry \cite{Sakharov:1967dj}. Since photons are neutral (i.e. they are their own antiparticles), 
the role played by particle number in baryogenesis must here be replaced by the polarization state 
of the radiation field. Inside a region with a specified vacuum and no electromagnetic sources, 
different polarization states propagate independently and preserve 
their identity. Rebalancing among polarization states occurs only in the presence of charged matter, 
through Thomson scattering and related interactions, which couple the radiation field to the plasma. 
It is this coupling that allows cosmological perturbations to imprint a nontrivial 
polarization pattern on the CMB.

If an unpolarized photon bath was propagating through a perfectly homogeneous and isotropic 
charged plasma, no polarization pattern would be generated. The crucial ingredient is the 
presence of gravitational perturbations inherited from inflation. As recombination proceeds 
and the universe becomes increasingly transparent, the residual free electrons are redistributed 
by these perturbations, accumulating in some regions and becoming diluted in others. 
The resulting local quadrupole anisotropies in the radiation field generate linear polarization in spatial patches 
through Thomson scattering. The polarization orientation is determined by the 
local quadrupole distortion and is therefore correlated with the temperature 
anisotropies produced by the same perturbations.

The emergence of this polarized CMB is a dynamical process that requires both an evolving 
background and a departure from exact homogeneity and isotropy. Once generated, the 
polarization pattern is subsequently frozen in as the plasma dilutes and the universe expands. 
From this perspective, the production of a polarized CMB is analogous to a Sakharov-type 
asymmetry generating process \cite{Sakharov:1967dj}: a nontrivial polarization state is 
dynamically created and then preserved by cosmological evolution. By itself this does not 
induce birefringence, since the local interactions responsible for generating the polarization 
remain parity preserving. Up to this stage the role of the evolving background is only to 
generate and preserve the polarization pattern that later parity-violating dynamics can act upon.

Birefringence is generated at the next stage, when the sector that distinguishes the two 
photon helicities is activated. This is precisely the moment when thin domain walls 
(or, alternatively, ultralight axions) enter the scene. Sticking with the walls, this occurs when 
the corresponding Hubble parameter $H$ and bubble nucleation rate satisfy $\Gamma \ga H^4$. 
The nucleation process is then prolific and rapid \cite{Guth:1982pn,Turner:1992tz}, and as described 
in the previous section and in \cite{Kaloper:2025goq,Kaloper:2025wgn,Kaloper:2025upu}, the 
bubbles expand, collide and melt away into very light dark-sector states. The important point
is that this takes place after recombination. Consequently, the CMB photons, which already 
carry the polarization pattern generated by cosmological perturbations and Thomson scattering, 
must subsequently cross the domain walls. As long as the last scattering surface lies in the 
initial vacuum, prepared by a combination of inflation and local dark-sector vacuum dynamics, 
while the late universe resides in a final vacuum, the polarization of the CMB is rotated by 
the same amount along every line of sight, as explained in the previous section. 

Let us now illustrate this discussion using a sequence of simple steps which sketch the 
underlying physical picture. Imagine first that the last scattering surface is completely 
homogeneous, without any perturbations, and insert, for simplicity only, a single thin wall 
bearing Chern--Simons terms between it and us. In this case the photons propagating 
from last scattering would be completely unpolarized, and the wall would produce no 
observable effect despite distinguishing the two photon helicities. The reason is simple: 
in the absence of a primordial polarization pattern, there is nothing for the wall to act upon.

Further, even if the incoming CMB were polarized by some external means, a 
wall-induced rotation would still be impossible to identify without an independent 
reference pattern. One could observe a particular polarization orientation, but there 
would be no way to determine whether it was primordial or the result of a rotation 
acquired during propagation. In other words, the effect could be absorbed into a 
redefinition of the polarization basis or a reorientation of the instrument. In this 
idealized setup we could rotate the polarization of the CMB at will and 
remain none the wiser about the presence of the wall.

In the physically relevant case, however, the last scattering surface and the radiation 
evolving beyond it are {\it not} completely homogeneous. They are perturbed, giving 
rise to correlations between the CMB temperature anisotropies and the polarization 
field, which provide the reference pattern absent in our example. In this case, a wall 
rotates the polarization of the CMB relative to its primordial orientation, altering its 
correlation with the temperature anisotropies. 

The necessity of temperature fluctuations 
for observable birefringence was already noted for ultralight-axion scenarios 
in \cite{Lue:1998mq}, and it remains equally necessary here.
Only when perturbations generate both the temperature anisotropies and the linearly 
polarized CMB field can observations reveal the presence of the wall. 
The perturbations create the polarization pattern, while the parity-violating sector 
subsequently acts on it. The observable birefringence signal arises from 
the interplay of these two ingredients.

Turning to the details of the signature of a wall which twists a 
CMB field, as explained in \cite{Lue:1998mq,Minami:2020odp},
the presence of an optically active region along the CMB trajectory leads to the emergence
of components in the stochastic CMB field which parameterize 
T and P violation, $C_{\ell}^{TB}$ and $C_{\ell}^{EB}$. They come
from the rotation of the original polarization induced during the epoch of last scattering. It suffices to focus on the
moments of the $C_{\ell}^{EB}$ variable, 
\be
C_{\ell}^{EB} = \frac{1}{2 \ell + 1} \sum_{m = -\ell}^{\ell}
\frac{1}{2} \Bigl( E_{\ell m} B_{\ell m}^\ast + E_{\ell m}^\ast B_{\ell m} \Bigr) \, .
\label{cmoms}
\ee
and evaluate them using Boltzmann equation solvers which produce the 
terminal observables at our world point \cite{Nakatsuka:2022epj}. 

Conveniently, \cite{Nakatsuka:2022epj} begin
their discussion by considering what they refer to as ``toy examples" where they consider the polarization
rotation to be given by a step function on redshift after last scattering. It should be clear from the discussion
in the previous section and the depiction of cosmology in Fig. \ref{fig1} that this is {\it precisely} a simplified model of thin
domain wall-induced birefringence. It is the limit where we take the percolation of the terminal vacuum to be very 
fast, at most set by the Hubble time at the onset of nucleations. This is certainly a consistent limit given that
we consider the bubbles with nucleation rate $\Gamma \ga H^4$ then. So using 
$\vartheta = \Delta \theta [1-\Theta(z-z_{PT})]$, where $z_{PT}$ denotes the redshift of the phase transition
between the initial and terminal vacua, we obtain the spectra for $C_{\ell}^{EB}$, and as a benchmark 
also spectra for $C_{\ell}^{EE}$ using numerical integration. 

\begin{figure}[thb]
    \centering
    \includegraphics[scale=0.59]{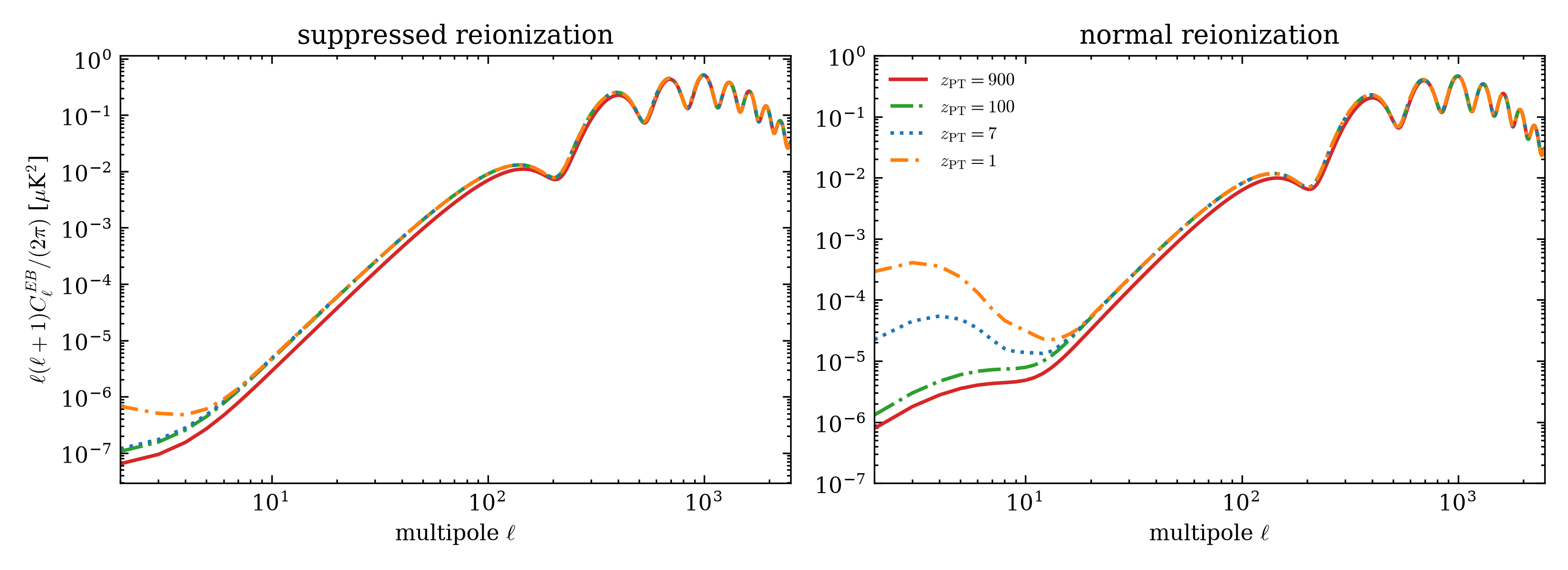}
    \caption{Spectra of $C_{\ell}^{EB}$ for transitions 
    at different redshifts $z_{\rm PT}$. Left panel: suppressed reionization; right panel: normal reionization.}
    \label{fig:CEB}
\end{figure}

\begin{figure}[thb]
    \centering
    \includegraphics[scale=0.59]{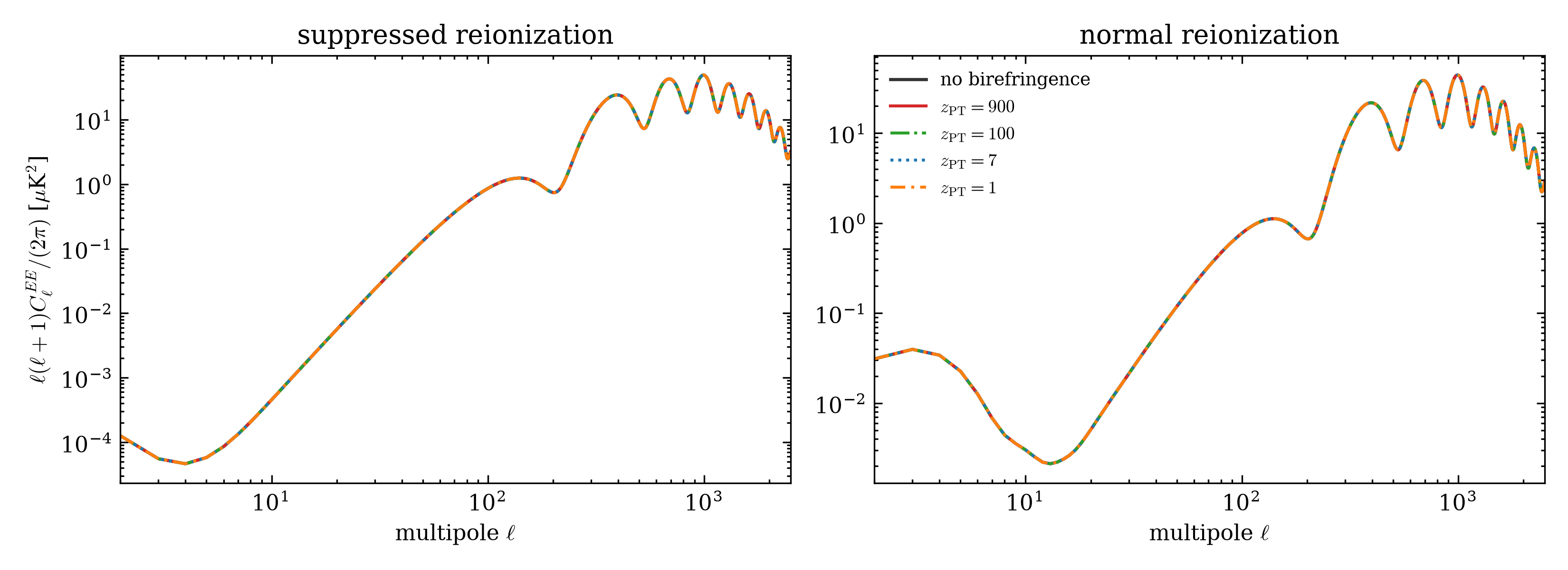}
    \caption{Spectra of $C_{\ell}^{EE}$ for the same transitions as in Fig. \ref{fig:CEB}.}
    \label{fig:CEE}
\end{figure}

In these plots, we have included for comparison both the late evolution of the signals with reionization due to
structure formation suppressed (left panels) and unsuppressed (right panels). In our case the presence of the reionization
bump might only affect the polarization rotation signal when the phase transition which sources the bubbles with 
thin walls bearing Chern--Simons terms is very late. The reason is that reionization induces a free electron plasma,
and so incoming light undergoes a second stage of Thomson scattering, which may shift the polarization of
light passing through it. 
If the phase transition occurs early, the wall-induced polarization rotation will be already
imprinted into the CMB field, and will not be affected significantly at most wavelengths. We indeed see
this in the plots in Figs.~\ref{fig:CEB} and \ref{fig:CEE},  since 
high-$l$ modes correspond to short distances, that are directly affected by scattering 
on domain walls. Further, we also see that the polarization rotation contributions from reionization bump
are barely near the current observational thresholds only when the phase transition occurs around, or after
reionization. Essentially, in this case the reionization Thomson scattering slightly shifts the incoming CMB polarization
from its value set by last scattering, and then it is the shifted fields which are rotated by the walls.

These effects are different from what happens when one considers cosmologically evolving ultralights. 
As we explained above the thin wall-induced cosmic birefringence signal is isotropic in stark contrast to
what happens with ultralight fields undergoing cosmological evolution \cite{Namikawa:2024sax}. Thus
constraints on anisotropies may be a tool to discriminate between the two birefringence sources \cite{Lonappan:2025hwz}. 

Secondly, the wall-induced and ultralight-induced birefringence lead to very different redshift dependence of the
observables. The redshift that corresponds to when the polarization rotation is imprinted might be determined by 
the tomographic approach if one observes the effects of the reionization bump at $\ell \lesssim 20$ in $C_{\ell}^{EB}$, 
expected if nonzero polarization rotation occurs after reionization \cite{Sherwin:2021vgb,Nakatsuka:2022epj}.
Clearly, this could be a potent tool for discriminating between various types of ultralight-induced birefringence. 
In the case of thin domain walls which we are mainly interested in, as our Figs.~\ref{fig:CEB} and \ref{fig:CEE} show this 
approach might only be viable if the domain wall-induced birefringence occurs at very low redshifts.

Another specific signature of thin domain wall birefringence 
may come from astrophysical sources. In general, this is challenging at present since 
astrophysical probes which could be used -- called \emph{standard crosses} \cite{Naokawa:2025shr} -- 
are less ``taxonomized" at present. While the CMB is a stochastic field with well-characterized 
properties, standard crosses emit polarized radiation whose intrinsic polarization direction 
is only approximately known, and one must infer a rotation 
relative to that intrinsic direction. 

A promising subset of sources 
are extended radio galaxies, whose synchrotron emission can be 
observed. The polarization rotation can then be measured using, for 
instance, the methods proposed in \cite{Whittaker:2017hnz} and \cite{Yin:2024fez}.
Upcoming large radio surveys turn these noisy sources into 
statistical ``standard crosses" for measuring birefringence.
As pointed out in \cite{Naokawa:2025shr}, a sample of about 
$10^6$ galaxies, each serving as an imprecise standard cross, 
can detect a rotation angle $\Delta \theta \simeq 3 \times 10^{-3}$ radians, even 
assuming a scatter of order a radian around the mean intrinsic polarization angle.
This sample size is not unrealistic in the medium term: 
for reference, \cite{OSullivan:2023eub} currently reports a 
sample of 2500 objects observed up to $z=2$.

The sharpest signal that would discriminate between thin walls as a source of 
birefringence versus ultralight fields is that even if a 
CMB signal were observed, radio galaxies were to show no sign of polarization rotation -- a null signal. The radio galaxies 
emit in the electromagnetic spectrum around or below CMB frequencies. The positive signal of birefringence in the
CMB and the absence of such a signal in the radio galaxies sample would indicate that the mechanism behind birefringence
is the walls which were produced early, at redshifts before the radio galaxies were formed. This kind of step function
would be a tell tale of walls which nucleated and percolated at redshifts before the ``first light" emerged. 

Another potentially measurable prediction of the walls bearing Chern-Simons terms might come from 
polarization rotation frequency dependence. The topological nature of thin wall birefringence implies that
below the cutoff of the theory all frequencies are affected the same. Above 
the cutoff of the field theory describing the
walls, the propagating cosmological light would not be affected significantly. 

\section{Summary}

We have argued that observations of cosmic birefringence may provide a rare opportunity
to confront ideas about dark energy, axions and quantum gravity with experiment.
This possibility is particularly intriguing because there is an observational hint of small 
polarization rotation $\sim 10^{-3}$ radians \cite{Komatsu:2022nvu}. Although small, 
explaining it appears to bring together several questions that are usually discussed independently.
If birefringence is real, one may need very light degrees of freedom, highly protected
effective field theories, large field excursions, nontrivial vacuum structure, or some
combination thereof. Thus, surprisingly, a phenomenon observed in the propagation of microwave photons
might be probing physics extending far beyond conventional particle phenomenology. 

Before summarizing our results, we need to highlight a cautionary remark which we view as
essential. It is important to clearly separate cosmological birefringence induced by 
slowly varying time-dependent ultralight axions from birefringence associated with axionic domain walls
or vacuum interfaces, thick or thin. The formulas describing polarization rotation may  
appear superficially similar, but the underlying physics is completely different.
In the case of a rolling ultralight field, the polarization rotation accumulates
adiabatically over cosmological distances and is controlled by the evolution and
initial conditions of a propagating degree of freedom. For domain walls, regardless of thickness, 
the rotation is tied to vacuum structure itself. For thick walls the effect is set
by the wall profile, which is ultimately determined by the topology of the vacuum
manifold; in this case the adiabatic approximation for computing birefringence effects still works
and this is why the thick walls are often conflated with cosmologically varying ultralight fields. 
However for thin walls and vacuum interfaces, adiabatic approximation fails, but the 
effect persists just fine. In this limit birefringence is controlled by discontinuous
changes of the effective vacuum angle, that yields a non-vanishing Pancharatnam phase 
\cite{Pancharatnam:1956url,Pancharatnam:1956url2}. Thus these mechanisms should not be identified
with one another merely because they can yield formally similar expressions
for polarization rotation.

The commonly invoked explanation of cosmic birefringence assumes an ultralight axion
coupled to electromagnetism through a Chern--Simons interaction. If a time-dependent dynamical 
axion is to affect the CMB after last scattering, it must
remain frozen until at least the epoch of last scattering, requiring masses as small as
$m \lesssim 10^{-28}\,\mathrm{eV}$ and often considerably smaller. Further, the 
axion decay constants are bounded by asking that the energy in the axion does not 
overclose the universe. If the axion is light enough to be dark energy, the mass drops additional
four orders of magnitude, and the required field excursions are trans-Planckian. 
Obtaining observable birefringence then requires that these ultralight fields retain
a sufficient coupling to electromagnetism, which is not a small feat.
Taken together, these ingredients point toward a highly special corner of effective
field theory.

In response we revisited the question of how an ultralight axion might
inherit an observable electromagnetic coupling. We reviewed simple see-saw
constructions and their extension to multi-axion sectors with ``Clockwork''-like
mass hierarchies. The principal lesson is that generating hierarchies of 
masses and couplings need not be the same phenomenon. The familiar suppression of couplings
in ``Clockwork" constructions relies on a specific localization of the coupling
vector in axion field space. If couplings are communicated before the hierarchy
is generated, a parametrically light mode may inherit a significantly larger
electromagnetic interaction than folklore sometimes suggests. Our purpose was to 
at least identify if such a region of parameter space might be realized in
effective field theory. We remain agnostic about whether Nature
actually realizes such a construction. 

The need for extremely light fields, large excursions and observable couplings
naturally leads to questions associated with the Weak Gravity Conjecture and
the broader Swampland program. It turns out that the observationally
interesting regime for ultralight birefringence tends to drift into territory
where these considerations become relevant. Using both the familiar electric version
of the axionic WGC and a magnetic formulation tied to the cutoff of the sector
generating the nonperturbative potential, we found evidence for a genuine
tension between ultralight birefringence and present Swampland intuition.
We deliberately refrain from claiming a theorem or an exclusion. The uncertainties
associated with numerical coefficients, ultraviolet completions and the detailed
implementation of the conjectures remain substantial and controversial. Nevertheless, the coincidence
is striking: the masses favored by birefringence and the mass bounds suggested by
Swampland reasoning appear uncomfortably close to one another, 
and trending in the opposite directions. If the interpretation
of birefringence in terms of ultralight rolling fields is ultimately confirmed,
some of our present expectations about large field excursions and quantum gravity
may require revision.

However, this is not the only game in town any more. Recent developments 
\cite{Kaloper:2026slg,Kaloper:2026gib,Kaloper:2026ygk} have uncovered a qualitatively 
different mechanism for birefringence which does not rely on ultralight propagating fields at all. 
In this framework, the relevant degrees of freedom are vacuum interfaces or thin domain walls 
carrying residual electromagnetic Chern--Simons couplings. Crossing such
interfaces induces a discrete polarization twist determined by the topology of
the vacuum configuration. The observed rotation angle can naturally be of order
$10^{-3}$ without invoking quintessence, Planckian field excursions or
propagating ultralight axions. These constructions fit naturally
within scenarios such as Discretely Evanescent Dark Energy \cite{Kaloper:2025goq}, 
where vacuum decay and its associated domain walls occur only recently on cosmological
timescales. 

Interestingly, the observable birefringence follows from dynamics that is a variant
of Sakharov's asymmetry generating process \cite{Sakharov:1967dj}. 
Cosmology first prepares a linearly polarized CMB, and a parity-violating sector later twists 
it into a measurable chiral signal. This closely follows the previously identified framework
explaining how ultralights can induce CMB birefringence \cite{Lue:1998mq}.

Very importantly, the ultralight-sourced and domain wall-sourced birefringence mechanisms make 
fundamentally different observational predictions. In the ultralight scenario, birefringence
is accumulated throughout the bulk of spacetime and is therefore sensitive to
the history, fluctuations and initial conditions of the underlying field.
Anisotropies are not an accident but a generic consequence. In contrast,
vacuum-interface birefringence is determined only by the initial and final vacua
encountered along a line of sight. Under the minimal assumptions discussed here,
the resulting signal is isotropic. The distinction is sharp. An anisotropic
signal points toward rolling ultralight fields. An isotropic signal points
toward vacuum interfaces. Future measurements by experiments such as
POLARBEAR, Simons Observatory and LiteBIRD might do considerably more
than establish the existence of birefringence. 
They might help identify the physical mechanism behind it, 
and arguably even end up probing for hints of quantum gravity.

Additional probes may strengthen this discrimination. If the actual CMB birefringence is 
eventually confirmed, tomographic information,
reionization signatures, measurements of anisotropic birefringence and
polarization studies of distant radio galaxies could all help determine when
and how the rotation is imprinted. In particular, a positive CMB signal combined
with the absence of a corresponding signal from later astrophysical sources
would strongly favor a vacuum-interface origin. Conversely, evidence for
direction-dependent polarization rotation would naturally support a cosmological
ultralight field, and likely bring a dark cloud over the swampland.

We therefore arrive at a position that is simultaneously cautious and optimistic.
The present data do not compel us to abandon either ultralight axions or
Swampland ideas. Nor do they require vacuum-interface birefringence.
What they offer instead is something far more valuable: a possible empirical
window into a domain of theory that has largely been guided by intuition and 
consistency arguments. If future observations reveal anisotropic birefringence
associated with cosmological ultralights, we may be forced to reconsider some
of our current expectations regarding field ranges, axions and quantum gravity.
If instead birefringence proves isotropic and consistent with vacuum interfaces,
much of the apparent tension may evaporate. Either outcome would teach us
something important. The point is not that the string bog has been mapped,
but that observations may finally provide some trails to navigate it.

\vskip.3cm

{\bf Acknowledgments}: AW is partially supported by the Deutsche
Forschungsgemeinschaft under Germanyâ Excellence Strategy - EXC 2121, ``Quantum
Universe\,IIâ" - 390833306 and by the
Deutsche Forschungsgemeinschaft through the Collaborative Research Center SFB1624, 
``Higher Structures, Moduli Spaces, and Integrability."

\end{document}